\documentclass[reprint,superscriptaddress,amssymb,amsmath,aps,showpacs,10pt,longbibliography,prb]{revtex4-2}
\usepackage[utf8]{inputenc}
\usepackage{textcomp}
\usepackage{graphicx}

\graphicspath{{figures/}}
\usepackage{color}
\usepackage[colorlinks,bookmarks=false,citecolor=darkblue,linkcolor=red,urlcolor=blue]{hyperref}

\definecolor{darkred}{rgb}{0.7,0.0,0.0}
\definecolor{darkblue}{rgb}{0,0.02,0.45}

\definecolor{darkgreen}{rgb}{0.02,0.45,0.0}

\begin{document}


\title{Quantum spin ladder with
ferromagnetic rungs in Bi$_2$CuO$_3$(SO$_4$)}

\author{Rodolfo A. Rangel Hernandez}
\affiliation{Felix Bloch Institute for Solid-State Physics, University of Leipzig, 04103 Leipzig, Germany}

\author{Kirill~Yu.~Povarov}
\affiliation{Dresden High Magnetic Field Laboratory (HLD-EMFL) and W\"urzburg-Dresden Cluster of Excellence ctd.qmat, Helmholtz-Zentrum Dresden-Rossendorf (HZDR), 01328 Dresden, Germany}

\author{Sergei Zvyagin}
\affiliation{Dresden High Magnetic Field Laboratory (HLD-EMFL) and W\"urzburg-Dresden Cluster of Excellence ctd.qmat, Helmholtz-Zentrum Dresden-Rossendorf (HZDR), 01328 Dresden, Germany}

\author{Oleg I. Siidra}
\affiliation{Department of Crystallography, St. Petersburg State
University, 199034 St. Petersburg, Russia}

\author{Alexander A. Tsirlin}
\email{altsirlin@gmail.com}
\affiliation{Felix Bloch Institute for Solid-State Physics, University of Leipzig, 04103 Leipzig, Germany}

\author{Victoria A. Ginga}
\email{victoria.ginga@uni-leipzig.de}
\affiliation{Felix Bloch Institute for Solid-State Physics, University of Leipzig, 04103 Leipzig, Germany}


\begin{abstract}
We introduce Bi$_2$CuO$_3$(SO$_4$) as a rare
example of a spin-ladder magnet with ferromagnetic interactions on the rungs. Its magnetic response is studied through measurements of heat capacity, temperature-dependent magnetic susceptibility, and field-dependent magnetization, as well as electron spin resonance spectroscopy. These experiments are complemented by density-functional-theory calculations combined with the construction of maximally localized Wannier functions and an analysis of the relevant superexchange pathways. Quantum Monte Carlo simulations are employed to model thermodynamic properties and to quantitatively determine the magnetic exchange parameters. Our combined approach identifies Bi$_2$CuO$_3$(SO$_4$) as a two-leg spin-ladder system with ferromagnetic rungs ($J'$ $\approx -208$ K) and antiferromagnetic legs ($J$ $\approx 258$ K). These interactions of similar magnitude arise from remarkably different superexchange pathways, with the Cu--Cu distance along the leg being almost twice as long than the respective distance along the rung. 
The antiferromagnetic leg coupling represents the strongest oxygen-mediated long-range superexchange in a Cu$^{2+}$ compound reported to date and sets the benchmark for the role of complex superexchange pathways in quantum magnets.   
\end{abstract}

\maketitle


\section{Introduction}
Low-dimensional quantum magnets have been the subject of extensive theoretical and experimental interest for several decades. These materials display a variety of unconventional ground- and excited-state properties that emerge from strong quantum fluctuations and competing interactions ~\cite{haldane1983,dagotto1996, inosov2018,zhang2020}. Transition-metal compounds containing Cu$^{2+}$ ions with a 3$d^9$ configuration provide numerous realizations of such physics, since subtle changes in the local coordination environment can drastically modify the effective magnetic dimensionality and stabilize diverse magnetic ground states ~\cite{inosov2018,lemmens2003,lebernegg2017,kulbakov2022a}. An effective strategy to tune the magnetic dimensionality in Cu-based systems involves the incorporation of stereochemically active lone-pair cations such as Bi$^{3+}$ or Te$^{4+}$ ~\cite{glamazda2017,tsirlin2010}. Their 6$s^2$ (or 5$s^2$) lone pairs introduce asymmetric local distortions that modify Cu–O–Cu bond geometries and thereby reshape the underlying superexchange network.

Several of the copper-based $S$ = 1/2 materials adopt the spin-ladder geometry. The Cu$^{2+}$ ions form networks of corner- or edge-sharing CuO$_4$ plaquettes that further assemble into two-leg or multi-leg ladders, resulting in markedly different magnetic ground states. Even-leg ladders feature a finite spin gap, whereas odd-leg ladders behave similarly to single Heisenberg chains and feature gapless spin excitations ~\cite{dagotto1996,azuma1994,glamazda2017}. 
One common scenario of the ladder formation is the condensation of CuO$_4$ plaquettes via single oxygen atoms, as in SrCu$_2$O$_3$ ~\cite{azuma1994}, CaCu$_2$O$_3$~\cite{kiryukhin2001}, and La$_2$Cu$_2$O$_5$~\cite{sugai1999} whose magnetic behavior is well described by the model of a two-leg spin ladder with dominant rung–leg interactions mediated by Cu--O--Cu superexchange, and no competing exchange pathways.

\begin{figure*}[t]
  \centering
  \includegraphics[width=1\textwidth]{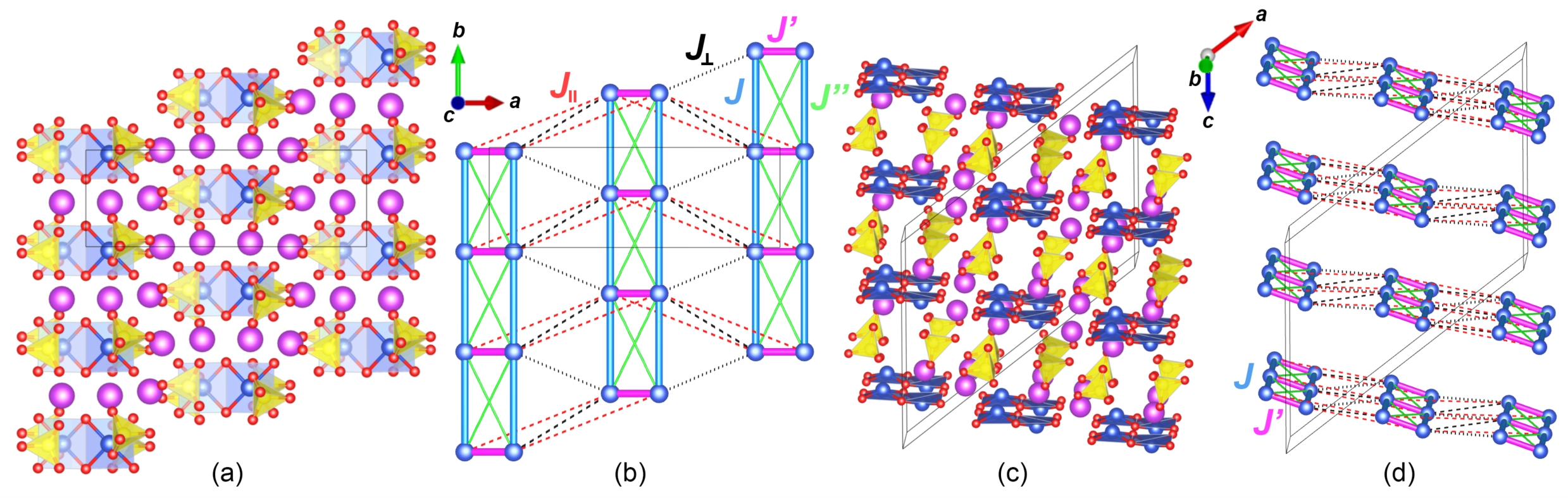}
    \caption{General projections of the crystal structure and spin lattice of Bi$_2$CuO$_3$(SO$_4$) along $c$ (a,b) and within the $ac$ plane (c,d) (Cu - blue balls, Bi - purple balls, CuO$_4$ plaquettes are shown in blue, (SO$_4$)$^{2-}$ tetrahedra are yellow). The spin lattice comprises the leg $J$ (cyan), rung $J'$ (pink), diagonal $J''$ (green), $J_{\parallel}$ (red) and $J_{\perp}$ (black) couplings. $J_{\perp}$ interactions are illustrated with different line styles: the 5.68\,\r A pathway is shown as a black dotted line, whereas the 6.09\,\r A pathway is shown as a black dashed line.}
    \label{fig1}
\end{figure*}

In contrast, BiCu$_2$PO$_6$~\cite{koteswararao2007}, which crystallizes in the orthorhombic space group \textit{Pnma} and comprises ribbons of edge-sharing CuO$_4$ plaquettes forming coupled two-leg ladders along the $b$ axis, stands out as one of the few known $S$ = 1/2 ladder systems with magnetic frustration caused by sizable antiferromagnetic (AFM) next-nearest-neighbor (NNN) exchanges along the ladder legs~\cite{mentre2009,tsirlin2010}. These NNN pathways compete with nearest-neighbor interactions, which are also AFM, and give rise to a peculiar energy spectrum that features singlet bound modes~\cite{choi2013} and the quasiparticle-continuum level repulsion~\cite{plumb2016}, along with a series of field-induced phases~\cite{kohama2012}, including the quantum soliton lattice~\cite{casola2013}. Magnetic anisotropy in the form of Dzyaloshinskii–Moriya interactions is known to be important for the quantitative description of this material~\cite{splinter2016,hwang2016,pilch2025}. On the structural level, one of its peculiarities is the importance of long-range superexchange via two consecutive oxygen atoms that underlies the rung couplings as well as the NNN couplings along the legs~\cite{tsirlin2010}.

In the following, we introduce another member of the family of BiCu-based spin-ladder materials, the previously unexplored Bi$_2$CuO$_3$(SO$_4$)~\cite{siidra2020}. The compound Bi$_2$CoO$_3$(SO$_4$), although sharing the same nominal stoichiometry, adopts a distinct crystal structure featuring corner-sharing CoO$_5$ pyramids. The latter form antiferromagnetic chains with a weak ferromagnetic component in the ordered state below approximately 20 K attributed to spin canting~\cite{lu2014}. The Bi$_2$CuO$_3$(SO$_4$) compound turns out to show a very different magnetic behavior. Our experimental characterization of the magnetic response of Bi$_2$CuO$_3$(SO$_4$) combines measurements of the heat capacity, temperature-dependent magnetic susceptibility, and field-dependent magnetization, along with electron spin resonance spectroscopy. Complementing these measurements, we carry out density functional theory (DFT) band-structure calculations, construct Wannier functions. This allows us to determine the relevant exchange pathways through a mapping procedure. Quantum Monte Carlo simulations are then employed to model thermodynamic properties and quantitatively extract the underlying magnetic couplings. We show that Bi$_2$CuO$_3$(SO$_4$) is well described by a model of the two-leg spin ladder with AFM legs and ferromagnetic (FM) rungs (Fig.~\ref{fig1}). Long-range superexchange plays crucial role in the formation of spin ladders with the remarkably strong exchange couplings in excess of 200\,K despite large separations between the Cu$^{2+}$ ions.


\section{Methods}

Previously, Bi$_2$CuO$_3$(SO$_4$) was prepared by a high-pressure high-temperature reaction of CuSO$_4$ and BiOCl~\cite{siidra2020}. Polycrystalline samples of Bi$_2$CuO$_3$(SO$_4$) were prepared from CuSO$_4$ and BiOCl precursors mixed in a molar ratio of 1:2.2. The thoroughly ground precursor mixture were sealed under vacuum in quartz ampoules and heated following the same temperature profile (including heating and cooling rates) as described in Ref.~\cite{siidra2020}, but without applying external pressure. Two independent synthesis batches of Bi$_2$CuO$_3$(SO$_4$) were prepared and investigated. Their crystal structure and phase composition were verified by collecting high-resolution x-ray diffraction (XRD) data at 80\,K and 298\,K~\cite{esrf} at the ID22 beamline of the European Synchrotron Radiation Facility (ESRF) in Grenoble, France using the wavelength of 0.35433\,\r A and the multi-analyzer detector setup~\cite{fitch2023}. The crystal structure was refined by the Rietveld method using the \texttt{Jana2006} software~\cite{jana2006}. 

\begin{figure}[h]
  \centering
  \includegraphics[width=0.45\textwidth]{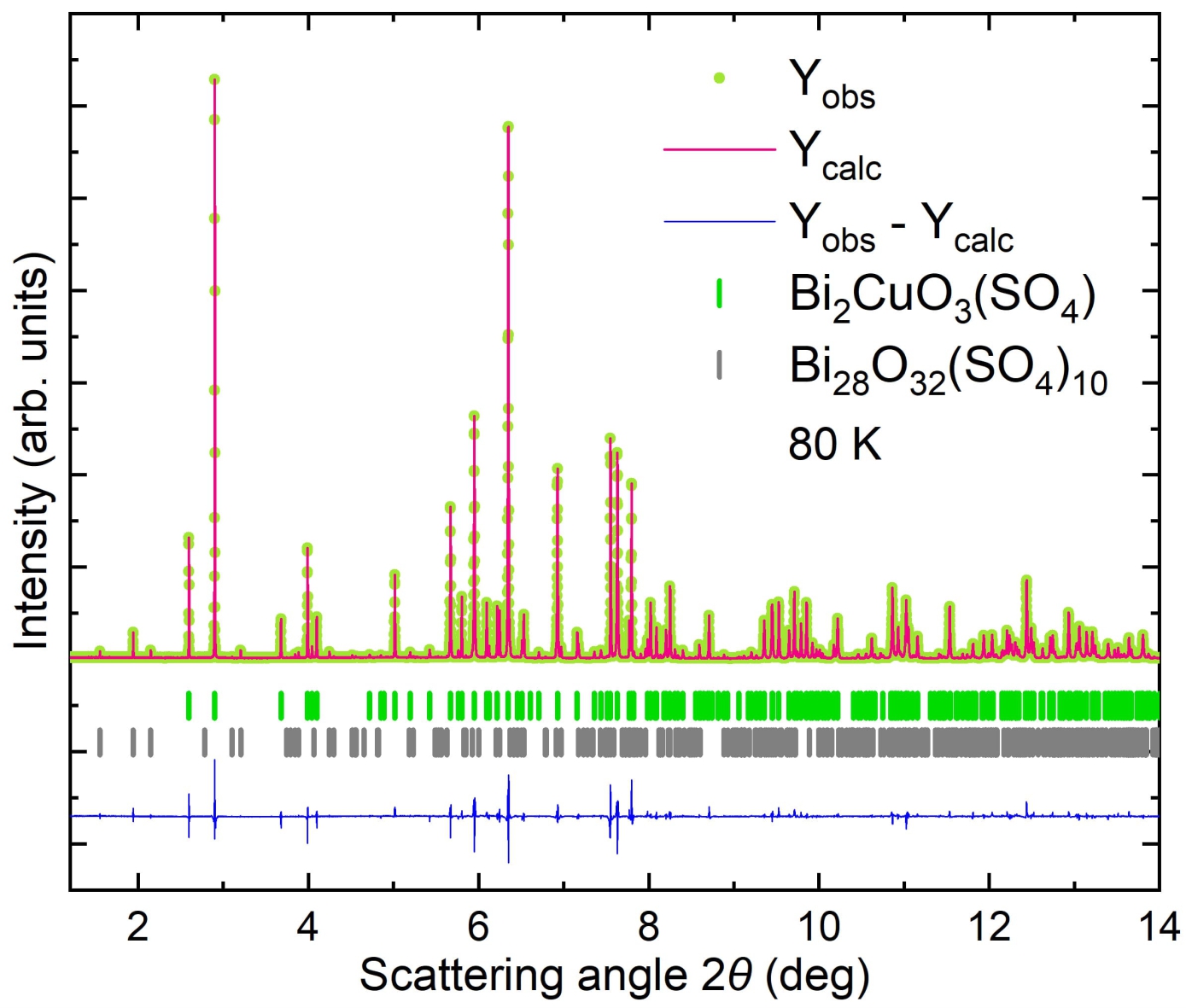}
    \caption{XRD pattern refinement of Bi$_2$CuO$_3$(SO$_4$) in the $C2/c$ space group by Rietveld method from the synchrotron powder data at 80 K (ID22, ESRF). Experimental data are shown as green circles (Y$_{obs}$), the calculated profile is shown by the red solid line (Y$_{calc}$) and the difference curve (Y$_{obs}$ - Y$_{calc}$) is displayed as a blue line at the bottom. Vertical tick marks indicate the Bragg reflection positions of the main phase Bi$_2$CuO$_3$(SO$_4$) (green) and the impurity phase Bi$_{28}$O$_{32}$(SO$_4$)$_{10}$ (gray).}
    \label{fig2}
\end{figure}

Rietveld refinements against the high-resolution XRD data (Fig.~\ref{fig2}) consistently reveal the presence of 8-10~wt.\% of the nonmagnetic impurity phase Bi$_{28}$O$_{32}$(SO$_4$)$_{10}$~\cite{aurivillius1987}. All measurements shown below have been performed on the sample that did not contain any further impurity phases. The sample from the second batch additionally showed several very weak reflections that could be tentatively assigned to the unreacted BiOCl and to Cu(OH)Cl, which might form as a secondary phase during the synthesis. While the latter impurity is magnetic, we did not observe any additional anomalies in the magnetization data. All features reported in this work are well reproducible across the two batches~\cite{SM} and should be intrinsic to Bi$_2$CuO$_3$(SO$_4$).


Magnetic susceptibility $\chi$(T) was measured on an 80.5 mg polycrystalline sample of Bi$_2$CuO$_3$(SO$_4$). The data were collected in applied magnetic fields of 0.1 T, 1 T, 2.5 T, and 5 T over the temperature range of 2–300\,K using a Quantum Design MPMS-XL magnetometer. For temperature-dependent magnetization measurements, both zero-field-cooled (ZFC) and field-cooled (FC) protocols were implemented. Field-dependent magnetization was additionally recorded in magnetic fields up to 7 T.

Heat-capacity $C_p$(T) measurements were carried out on a pressed pellet (10.1 mg) between 1.9 K and 200 K in external magnetic fields of 0 T, 1 T, 2.5 T, 5 T, and 9 T using the heat-capacity option of the Physical Property Measurement System (PPMS, Quantum Design).


High-frequency electron spin resonance (ESR) measurements were performed at the Dresden High Magnetic Field Laboratory (Hochfeld Magnetlabor-Dresden), using a tunable-frequency spectrometer (similar to the one described in Ref.~\cite{Zvyagin_PhysB_2004_ESRinCuGeO3}), equipped with a $16$~T superconducting magnet. We employed VDI microwave-chain sources (product of Virginia Diodes, Inc., USA) as radiation source ($100-500$~GHz frequency range), and a hot-electron n-InSb bolometer (product of QMC Instruments Ltd., UK), operated at 4.2 K, as a detector. We used 2,2-diphenyl-1-picrylhydrazyl (DPPH) as a standard frequency-field marker.

To quantify individual magnetic exchange interactions, scalar-relativistic DFT band-structure calculations were carried out using the \texttt{FPLO} code~\cite{fplo} and the Perdew–Burke–Ernzerhof (PBE) exchange–correlation potential~\cite{pbe96}. With PBE calculations, we identified relevant magnetic orbitals and evaluated hopping parameters $t_i$ via a fit with an effective one-orbital tight-binding (TB) model constructed from Wannier functions (WFs) centered on the Cu sites~\cite{eschrig2009}. The use of the WF approach allows for a well-defined evaluation of the hopping integrals $t_i$ and yields a transparent representation of the underlying magnetic orbitals. These hopping parameters were introduced into a Hubbard model with the effective on-site Coulomb repulsion potential $U_{\rm eff}$ = 8.5\,eV.  For half-filled bands in the strongly correlated regime ($t_i$ $\ll$ $U_{\rm eff}$), the Hubbard model can be further reduced to a Heisenberg model, yielding the antiferromagnetic contributions to the exchange couplings as $J_i^{\rm AFM}$ = 4$t_i^2$/$U_{\rm eff}$.

The ferromagnetic contribution to the superexchange interaction arises from mechanisms not captured within the one-orbital model used in the aforementioned analysis. In cuprate systems, such ferromagnetic exchange is commonly attributed to Hund's coupling on the ligand sites~\cite{mazurenko2007}. As an alternative approach, the magnetic exchange interactions were evaluated by explicitly accounting for strong electronic correlations within the mean-field DFT+$U$ method. Correlations in the Cu 3$d$ shell were treated with standard parameters $U_{\rm Cu}$ = 8.5 eV and $J_{\rm Cu}$ = 1 eV~\cite{mazurenko2014,bag2021}. The individual exchange constants $J_i$ of the Heisenberg Hamiltonian,
\begin{equation}
 \mathcal H=\sum_{\langle ij\rangle} J_{ij}\mathbf S_i\mathbf S_j
\end{equation}
where $S=1/2$ and the summation is over bonds, were obtained by a mapping procedure using total energies of four collinear spin configurations ~\cite{xiang2011}. In all calculations, experimentally determined lattice parameters and atomic positions were employed. Total energies were evaluated using a $k$-point mesh comprising up to 64 points in the first Brillouin zone.

Thermodynamic properties were further evaluated using quantum Monte Carlo (QMC) simulations performed with the \texttt{loop} algorithm~\cite{loop} of the ALPS simulation package ~\cite{albuquerque2007}. Simulations were done for finite lattices with periodic boundary conditions. Isolated spin ladder with the length of 
$L$ = 32 sites was used. This length is sufficient to eliminate finite-size effects for thermodynamic properties within the temperature range under investigation.

\begin{figure*}[!ht]
  \centering
  \includegraphics[width=1\textwidth]{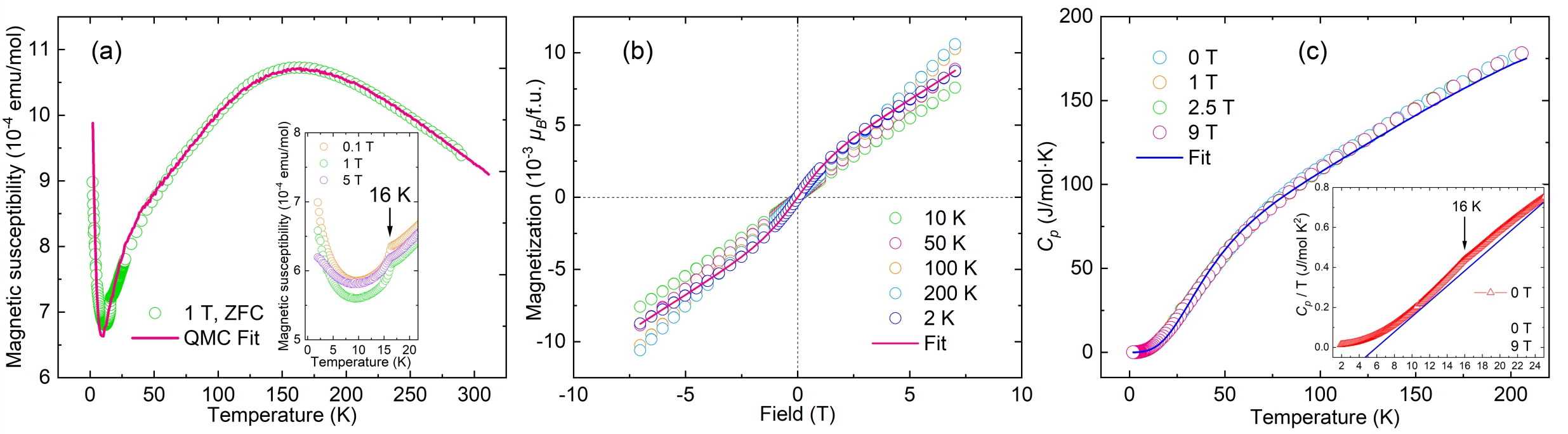}
    \caption{(a) Temperature-dependent magnetic susceptibility of Bi$_2$CuO$_3$(SO$_4$) measured in the applied field of 1 T with the QMC fit described in the text. The inset shows the low-temperature region of $\chi(T)$ at 0.1 T, 1 T, 5 T and highlights the anomaly at 16\,K. (b) Field-dependent magnetization of Bi$_2$CuO$_3$(SO$_4$) measured at 2 K, 10 K, 50 K, 100 K, and 200 K with the mean-field fit described in the text. (c) Temperature-dependent specific heat $C_p$(T) of Bi$_2$CuO$_3$(SO$_4$) measured at 0 T, 1 T, 2.5 T, and 9 T. The colored circles show the raw data, the blue line represents the estimation of phonon contribution using the model with two Debye temperatures, as explained in the text.}
    \label{fig3}
\end{figure*}


\section{Results}
\label{sec:results}
\subsection {Crystal structure}
High-resolution XRD data confirm the structural model of Bi$_2$CuO$_3$(SO$_4$) with the monoclinic space group $C2/c$~\cite{siidra2020}. No peak splittings or superstructure reflections could be observed at either 80\,K or 298\,K. The refined crystallographic parameters for Bi$_2$CuO$_3$(SO$_4$) in the $C2/c$ space group at 80 K and 298 K are summarized in Table~\ref{tab:lattice_parameters}. 

\begin{table}[h]
\caption{Refined crystallographic parameters for Bi$_2$CuO$_3$(SO$_4$) in the $C2/c$ space group at 80 K and 298 K.}
\label{tab:lattice_parameters}
\centering
\begin{tabular}{l@{\hspace{1cm}}c@{\hspace{1cm}}c}
\hline\hline
Parameter & 80 K & 298 K \\
\hline
$a$ (\AA) & 19.962(1) & 20.037(1) \\
$b$ (\AA) & 5.381(1) & 5.394(1) \\
$c$ (\AA) & 14.098(1) & 14.148(1) \\
$\beta$ ($^\circ$) & 128.389(1) & 128.397(1) \\
$V$ (\AA$^{3}$) & 1186.881(6) & 1198.467(6) \\
$R_1$ (\%) & 1.55 & 1.79 \\
\hline\hline
\end{tabular}
\end{table}

All lattice parameters increase slightly upon warming, indicating a moderate thermal expansion of the structure. A tentative estimate of the linear thermal-expansion coefficients returns $\alpha$$_a$ $\approx$ 18$\times$10$^{-6}$ K$^{-1}$, $\alpha$$_b$ $\approx$ 11$\times$10$^{-6}$ K$^{-1}$, and $\alpha$$_c$ $\approx$ 17$\times$10$^{-6}$ K$^{-1}$, indicating predominant thermal expansion within the $ac$ plane. This anisotropic thermal expansion highlights  quasi-one-dimensional (1D) nature of the material where spin-ladder units occur along the $b$ direction of the crystal structure (Fig.~\ref{fig1}).

The Cu site is surrounded by four oxygen atoms forming a distorted CuO$_4$ square with the Cu--O distances of $1.91-1.99$\,\r A, further capped by two apical O$^{2-}$ ligands with the longer distances of 2.26\,\r A and 2.76\,\r A, respectively~\cite{SM}. Together, they define a distorted [CuO$_4$O$_2$] octahedron. The equatorial planes CuO$_4$ of these [CuO$_4$O$_2$] octahedra are linked via edge sharing, effectively forming Cu$_2$O$_6$ dimers with the intradimer Cu–Cu distance of 2.79\,\r A. These dimers are arranged parallel to each other (Fig.~\ref{fig1}) and form spin ladders, as we show by the microscopic analysis in Sec.~\ref{sec:model}.

The dimers are linked via Bi$^{3+}$ ions that feature four shorter Bi--O bonds of $2.23-2.37$\,\r A and several further bonds that all exceed 2.65\,\r A. Such an asymmetry is typical for Bi$^{3+}$ with its stereoactive 6$s^2$ lone pair. One interesting feature of the material is that the shorter Bi--O bonds are always directed toward the CuO$_4$ plaquettes, whereas the longer bonds point toward the SO$_4$ tetrahedra. This leaves the tetrahedra rather weakly bonded between the planes formed by the CuO$_4$ plaquattes and Bi$^{3+}$ ions~\cite{SM}. Indeed, the refined crystal structure shows much higher atomic displacement parameters for the oxygen atoms of the SO$_4$ tetrahedra ($U_{\rm iso}=0.029$\,\r A$^2$) compared to the oxygen atoms of the CuO$_4$ plaquettes ($U_{\rm iso}=0.012$\,\r A$^2$) at 298\,K. Both values are reduced on cooling. At 80\,K, they become 0.009 and 0.005\,\r A$^2$, respectively, indicating that any disorder of the SO$_4$ tetrahedra is dynamic in nature.




\subsection {Magnetic susceptibility}

Temperature-dependent magnetic susceptibility $\chi$(T) of Bi$_2$CuO$_3$(SO$_4$) exhibits a broad maximum, characteristic of low-dimensional quantum spin systems (Fig.~\ref{fig3}a). The low-temperature upturn below 20 K is commonly associated with paramagnetic contributions from dilute impurities or defects, as further verified by ESR (Sec.~\ref{sec:esr}). The position of the maximum, $T_{max}$ $\approx 156$ K, shows only a weak field dependence and reflects the dominant exchange energy scale in the system. 
Magnetic susceptibility curves measured at different fields coincide, except in the region below 10\,K where the impurity contribution becomes prominent. The small anomaly observed in the magnetic susceptibility near 16 K is fully reproducible across different synthesis batches~\cite{SM} and likely reflects the onset of long-range magnetic order. We did not detect any magnetic impurity that could be responsible for this anomaly. 

The broad susceptibility maximum is typical for various low-dimensional spin systems. Therefore, we start analyzing the data using simple models with known analytical expressions for the magnetic susceptibility. The dimers of Cu atoms in the crystal structure (Fig.~\ref{fig1}) suggest a possible description in terms of spin dimers with 
\begin{equation}
\chi_{\rm dimer}(T) = \frac{N_A g^{2} \mu_{B}^{2}}{k_BT} \cdot \frac{1}{3 + e^{J/k_BT}}
\end{equation}
Alternatively, the spin-chain model can be used:
\begin{equation}
\chi_{\mathrm{chain}}(T) = 
\frac{N_A g^{2} \mu_{B}^{2}}{4 k_{B} T}
\,\frac{1 + \sum_{n=1}^{5} N_{n}/t^{n}}{1 + \sum_{n=1}^{6} D_{n}/t^{n}} 
\end{equation}
where $t$ = $k_{B}$$T/J$ and $N_{n}$ and $D_{n}$ are coefficients given in Ref.~\cite{johnston2000a}. However, none of these models reproduce the shape of the experimental susceptibility maximum, even if supplied with additional corrections for the temperature-independent and Curie-like impurity contributions (see Fig.~S3~\cite{SM}). In Sec.~\ref{sec:model}, we use a DFT-based microscopic magnetic model to fit the susceptibility maximum. Before presenting these results, we consider in more detail the nature of the low-temperature upturn that has to be taken into account when fitting the susceptibility data. 




\subsection {Magnetization}

Magnetization curves reveal no field-induced magnetic transition. For temperatures above 10 K, the magnetization exhibits a strictly linear dependence on the applied magnetic field, consistent with a predominantly paramagnetic or antiferromagnetic response. A deviation from the ideal paramagnetic behavior of $M(H)$ is visible at 2\,K, though (Fig.~\ref{fig3}b). This 2\,K magnetization curve is the only one that shows discernible Brillouin-type nonlinearity in the low-field range ($-2.5$\,T to 2.5\,T). As expected, such impurity-related contributions diminish rapidly with increasing temperature. 

To quantify this low-temperature behavior, we fitted $M(H)$ with a combination of the linear term and Brillouin function $B_S$,
\begin{equation}
M(B) = f \cdot N_{A} g \mu_{B} S \cdot B_{S}\!\left( \frac{S g \mu_{B} B}{k_{B} T} \right) + \chi_{m} \cdot B
\end{equation}
where $f$ stands for the percentage of impurities present in the sample. Using $S=\frac12$, one arrives at an unrealistically high $g$-value of 5.55 that indicates ferromagnetic correlations between the impurity spins. Such correlations can be taken into account on the mean-field level by introducing an effective field $B+\lambda M$ into the Brillouin function, 
\begin{equation}
M(B) = f\cdot N_Ag\mu_BS\cdot B_S\big(a(B+\lambda M)\big) +\chi_m\cdot B
\end{equation}
with $a = Sg\mu_B\,/\,k_BT$. This procedure results in a more realistic $g=2.10$ and $f=0.11\%$. 

The diminutively small amount of impurities inferred from this analysis leads to the susceptibility of about $\chi=2\times 10^{-4}$\,emu/mol at 2\,K, which is several times smaller than the experimental value of $9\times 10^{-4}$\,emu/mol. We thus conclude that the nonlinearity in $M(H)$ does not reflect the main impurity contribution, which arises from the impurity spins coupled antiferromagnetically. This finding is independently confirmed by ESR.


\subsection {Electron spin resonance}
\label{sec:esr}
\begin{figure}
\includegraphics[width=0.45\textwidth]{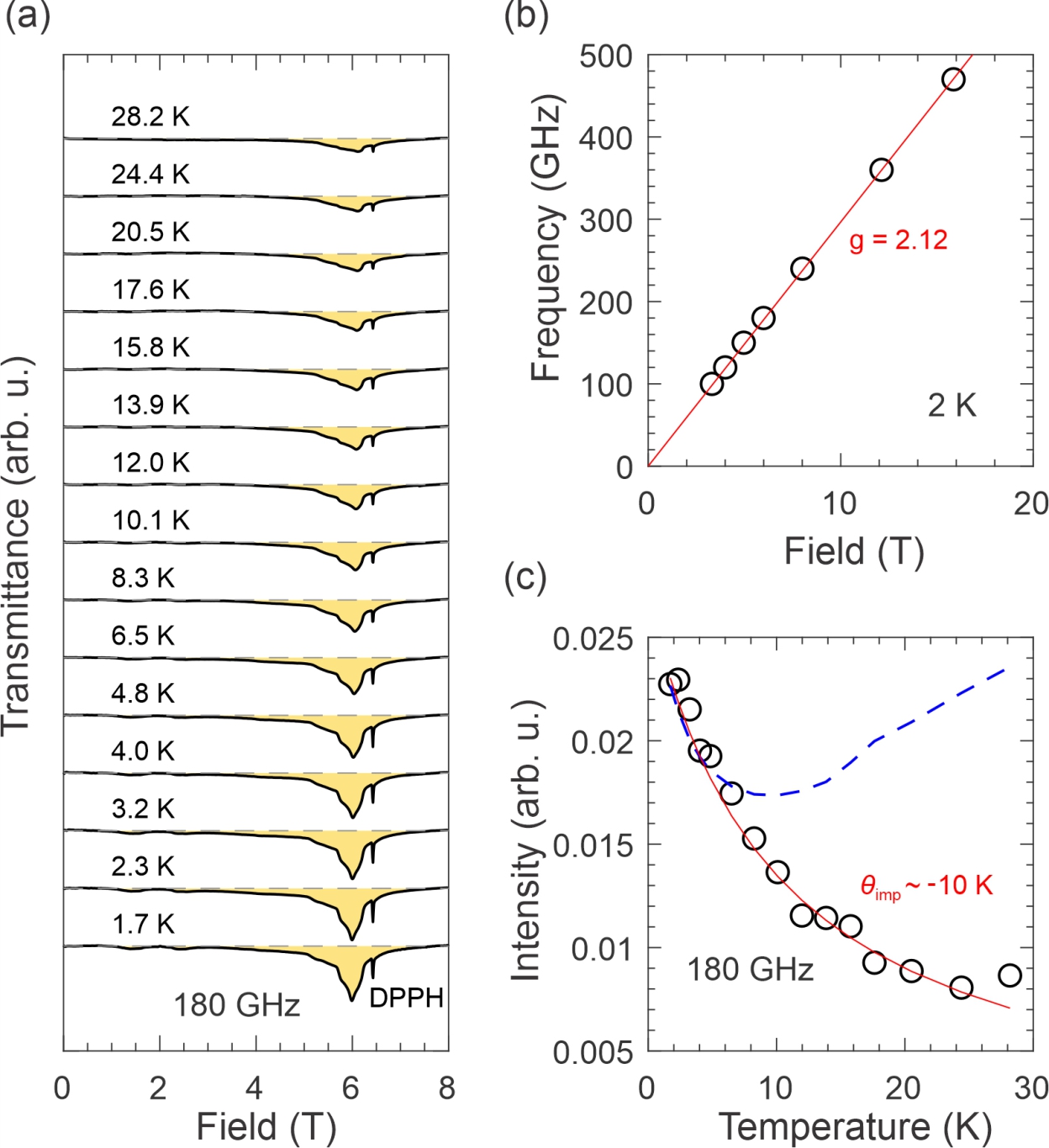}
\caption{(a) Exemplary ESR lines at different temperatures at $180$~GHz. Shading highlights the area under the curve. The sharp spikes in the spectra reflect the ESR signal of the DPPH marker. (b) Frequency-field diagram at $2$~K. Solid line corresponds to $g=2.12$. (c) ESR line intensity at $180$~GHz. Symbols represent the data, solid line is a Curie--Weiss fit. Dashed line is the bulk magnetic susceptibility scaled to match the ESR line intensity at low temperatures.}
\label{fig4}
\end{figure}

The exemplary ESR lines at $180$~GHz measured on a pressed pellet of Bi$_2$CuO$_3$(SO$_4$) are shown in Fig.~\ref{fig4}(a). The single observed transmittance dip has a relatively complicated profile, as typical for powder samples, and its amplitude is strongly temperature-dependent. The line at this frequency is centered around the resonant field of $6$~T at all temperatures. The low-temperature frequency-field diagram of the ESR response displayed in Fig.~\ref{fig4}(b) suggests that this resonance can be described with an effective powder-averaged $g$-factor of $2.12(2)$, which is a typical value for Cu$^{2+}$ ions.
For paramagnetic substances, the intensity of the ESR signal (corresponding to the area under the line) is proportional to the magnetization. Thus, it is instructive to compare the temperature-dependent ESR intensity to the bulk magnetic susceptibility. This comparison is shown in Fig.~\ref{fig4}(c) for the $180$~GHz data. The area estimate is done by numeric integration, with the DPPH contribution excluded. The observed ESR intensity clearly displays a Curie--Weiss behavior (solid line). This is a strong indication of the impurity subsystem being the source of the ESR absorption and the low-temperature upturn in $\chi(T)$. 

By describing the ESR intensity with a Curie-Weiss law, $\chi_{\rm imp}=C_{\rm imp}/(T-\theta_{\rm imp})$, we find $\theta_{\rm imp}\simeq -10$\,K and directly confirm that the dominant impurity contribution arises from the impurity spins coupled antiferromagnetically. They likely arise from structural imperfections that produce vacancies in the spin ladders of Bi$_2$CuO$_3$(SO$_4$). Such vacancies are known to behave as impurity spins with weak FM or AFM interactions depending on the separation between the impurities~\cite{sigrist1996,mikeska1997,schmidiger2016}.

\subsection {Heat capacity}

To verify the susceptibility anomaly at 16\,K, heat capacity of Bi$_2$CuO$_3$(SO$_4$) was measured in zero and applied magnetic fields. The resulting data are shown in Fig.~\ref{fig3}c. No field-induced effects are observed within the experimental resolution. Likewise, heating and cooling measurements reveal no substantial differences, indicating the absence of thermal hysteresis in the studied temperature range. 

The zero-field data were fitted in the range of 2-200 K using an empirical model with two Debye temperatures, 
\begin{equation}
C_p(T) =
9R \sum_{n=1}^{2} A_n
\left( \frac{T}{\theta_{D_n}} \right)^3
\int_{0}^{\theta_{D_n}/T} 
\frac{x^4 e^{x}}{(e^{x}-1)^2}\, dx 
\end{equation}
where $R$ is the universal gas constant, and $A_1$ and $A_2$ are dimensionless factors describing the relative weight of each Debye contribution to the total specific heat. The fit yields the parameters $A_1 = 6.59, \ \theta_{D_1} = 946.6~\mathrm{K}, \quad
A_2 = 4.41, \ \theta_{D_2} = 188.5~\mathrm{K}.$
While this fit does not account for the magnetic contribution, it highlights deviations from the simple Debye behavior. They are especially pronounced at $15-18$\,K where a kink is visible in $C_p/T$. This kink is reproducible between the different batches~\cite{SM}. It matches the temperature of the susceptibility anomaly and confirms its bulk nature. We thus conclude that Bi$_2$CuO$_3$(SO$_4$) undergoes magnetic ordering below 16\,K.

\subsection {Microscopic magnetic model}
\label{sec:model}

\begin{figure}[h]
\includegraphics[width=0.45\textwidth]{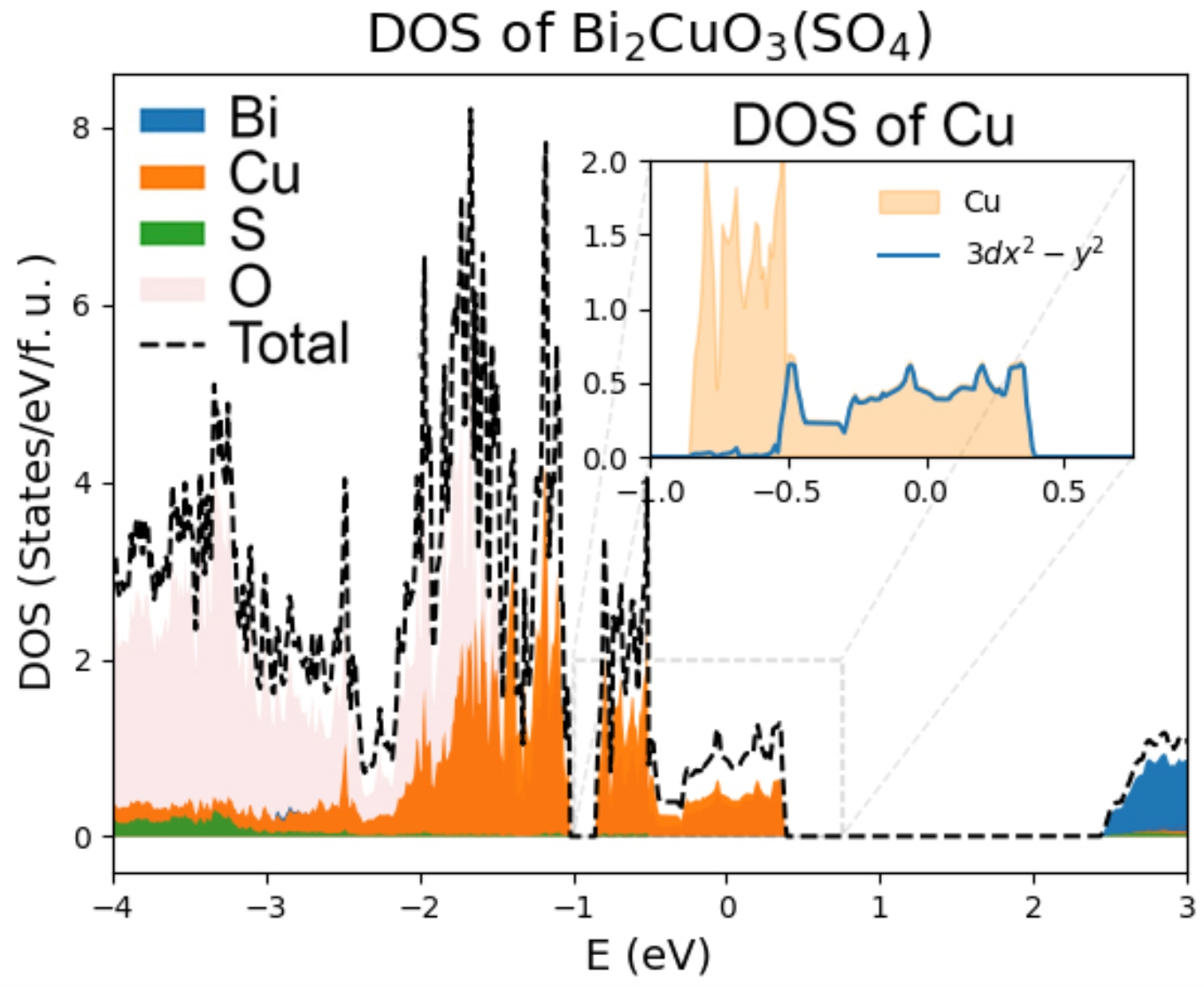}
\caption{Calculated electronic density of states for Bi$_2$CuO$_3$(SO$_4$) at the PBE level. Colored areas represent the element-resolved contributions of Bi, Cu, S, and O states, while the dashed line shows the total DOS. The Fermi level is set to $E = 0$ eV. The states in the vicinity of the Fermi level are dominated by Cu 3$d$ contributions. Inset: Cu-projected DOS highlighting the dominant contribution of the 3$d_{x^{2}-y^{2}}$ orbital.}
\label{fig5}
\end{figure}

PBE calculations for the band structure of Bi$_2$CuO$_3$(SO$_4$) are fully consistent with expectations for an insulating Cu$^{2+}$-based compound~\cite{tsirlin2010a,mazurenko2007,tsirlin2010b,tsirlin2011} (Fig.~\ref{fig5}). The valence band is predominantly composed of Cu 3$d$ and O 2$p$ states, while S 3$p$ orbitals contribute noticeably below $-3$ eV only. The states above $-0.6$ eV are dominated by the Cu 3$d_{x^{2}-y^{2}}$ orbital, in line with the anticipated ligand-field splitting. Unoccupied states above 2 eV arise from Bi 6$p$ orbitals. The partial densities of states clearly identify Cu$^{2+}$ ions as the magnetic ions of the material, and Cu 3$d$ bands additionally form narrow features close to the Fermi level. The metallic PBE spectrum contradicts the experimentally observed insulating behavior inferred from the green color of Bi$_2$CuO$_3$(SO$_4$). This discrepancy stems from strong correlation effects within the partially filled Cu 3$d$ shell that PBE significantly underestimates. These missing correlations can be incorporated either on the level of an effective model or within the mean-field DFT+$U$ approach.

\begin{figure}
\includegraphics[width=0.45\textwidth]{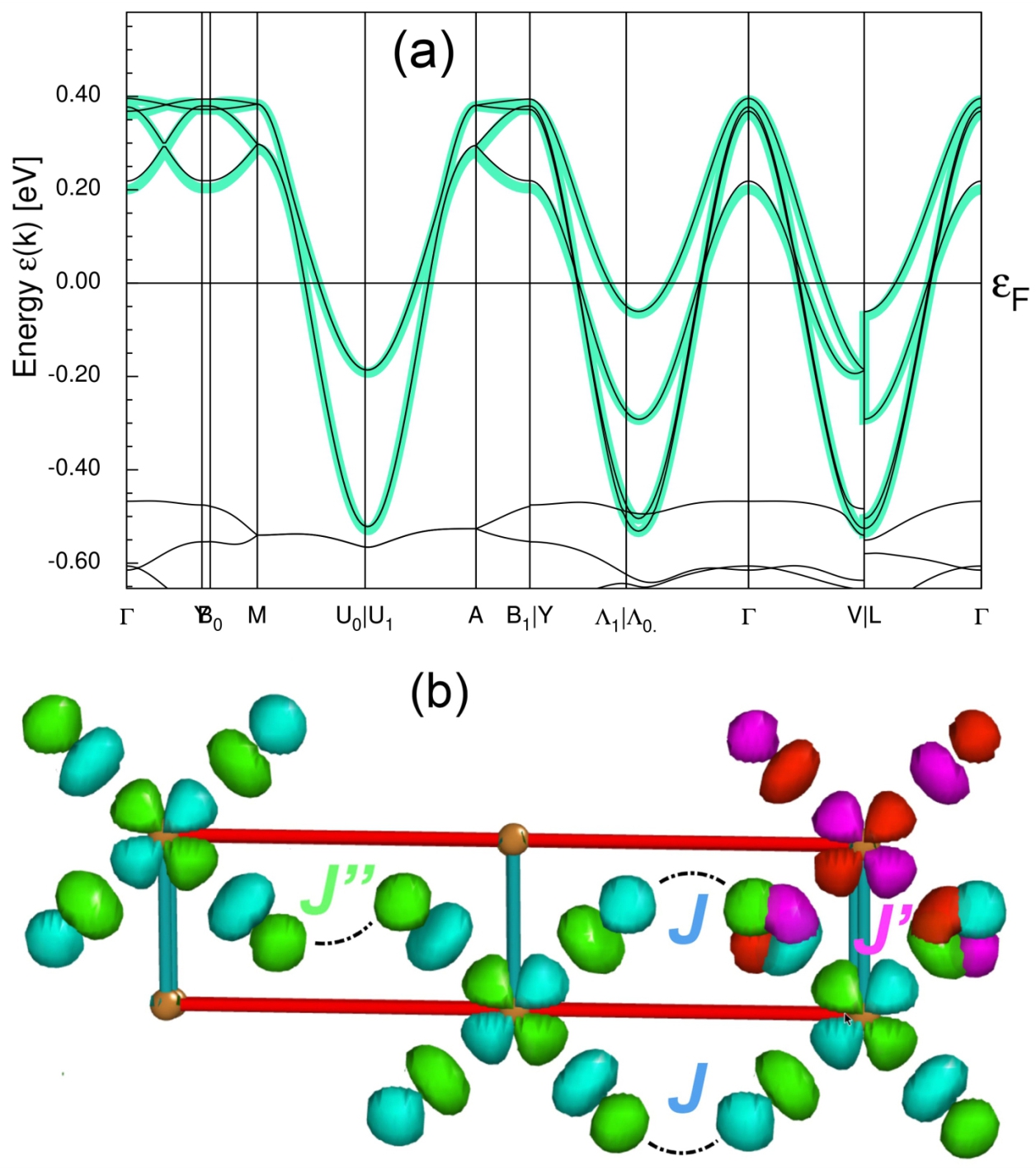}
\caption{(a) PBE band structure (thin black lines) and the fit with the tight-binding model (thick green lines).  (b) Magnetic orbitals represented through Wannier functions, where the Cu 3$d_{x^{2}-y^{2}}$ orbital exhibits pronounced hybridization with the surrounding O 2$p$ states. Oxygen atoms are omitted for clarity.}
\label{fig6}
\end{figure}

We focus on the narrow electronic bands intersecting the Fermi level. The bands are characterized by a dominant Cu 3$d_{x^{2}-y^{2}}$ orbital contribution, with the local orbital axes aligned along the shorter Cu–O bonds. In Bi$_2$CuO$_3$(SO$_4$), the Cu$^{2+}$ ions are situated in a strongly elongated octahedral coordination CuO$_{4+2}$, comprising four short in-plane Cu-O bonds (1.96–1.98\,\r A) and two significantly longer apical bonds (2.2\,\r A and 2.79\,\r A). This pronounced anisotropy of the local ligand environment places the 3$d_{x^{2}-y^{2}}$ orbital highest in energy within the crystal-field scheme, in full agreement with the PBE results. The 3$d_{x^{2}-y^{2}}$ -derived bands were subsequently described with an effective tight-binding model by constructing Wannier functions centered on the Cu sites~\cite{eschrig2009}. The resulting Wannier fit accurately reproduces the DFT band dispersion and allows for the extraction of the relevant Cu–Cu hopping integrals $t_i$ (Fig.~\ref{fig6} and Table~\ref{tab:exchange}). Additionally, we estimate the exchange couplings using the DFT+$U$ mapping approach that yields a sum of FM and AFM contributions.

The real-space Wannier functions are shown in Fig.~\ref{fig6}b and reveal a magnetic orbital of predominant Cu 3$d_{x^{2}-y^{2}}$ character, strongly hybridized with the surrounding O 2$p$ states that play a key role in mediating magnetic interactions. The Wannier representation provides direct insight into the exchange pathways along the spin ladder. The rung coupling $J'$ is governed by a Cu-O-Cu superexchange geometry with an angle of $90.11(1)^{\circ}$ close to $90^{\circ}$, which facilitates a FM interaction. In contrast, both the leg coupling $J$ and the diagonal coupling $J''$ are AFM and originate from extended Cu-O-O-Cu superexchange paths involving orbitals of several oxygen atoms. The drastic difference between $J$ and $J''$ is surprising in view of the same O--O contact and the very similar Cu--O--O angles of about $134^{\circ}$. An important geometrical difference is that the relevant $p$-orbitals of the oxygen atoms overlap in the case of $J$, whereas they are offset in the case of $J''$, thus hindering electron hopping and superexchange (Fig.~\ref{fig6}). 


\begin{table}[b]
\caption{\label{tab:exchange}
Cu--Cu distances (in\,\r A), hopping parameters $t_i$ (in meV),
and exchange couplings $J_i$ (in K) in Bi$_2$CuO$_3$(SO$_4$). The AFM contributions $J_i^{\rm AFM}$ are calculated as 4$t_i^2$/$U_{\rm eff}$ with $U_{\rm eff}$ = 8.5 eV. The total exchange couplings $J_i$ are obtained from DFT+$U$ calculations ($U_{\rm Cu}$ = 8.5 eV, $J_{\rm Cu}$ = 1 eV). The notation of $J_i$ is illustrated in Fig.~\ref{fig1}. The structural data reported in Ref~\cite{siidra2020} were used as input for the DFT calculations.
}
 
\begin{ruledtabular}
\begin{tabular}{ccccc}
coupling & $d_{\rm Cu-Cu}$ & $t_i$ & $J_i^{\rm AFM}$ & $J_i$ \medskip \\
 \hline \medskip
rung $J'$            & 2.79 & 62    &    20.6  & $-133.6$ \medskip \\
leg $J$             & 5.40  & 170   &  157.9   & 194.7 \medskip \\
diag $J''$           & 6.07 & $-49$ & 13.2   & 6.1  \medskip \\
$J_{\perp}$     & 5.68 &      8     &  0.3  &$-2.9$ \\
                & 6.09 &     $-41$  &  9.4  & 6 \\
 \medskip \\
$J_{\parallel}$ & 8.20 &     $-0.2$  &  $3 \times 10^{-4}$  & 1.8\\ 
                & 8.49 &     $-23$  &  2.9  & 1.8\\
\end{tabular}
\end{ruledtabular}
\end{table}

Based on the calculated exchange parameters summarized in Table~\ref{tab:exchange}, we conclude that the spin lattice of Bi$_2$CuO$_3$(SO$_4$) comprises non-frustrated spin ladders with FM rungs and AFM legs. The diagonal coupling $J''$ is smaller than $J$ by more than one order of magnitude and, therefore, negligible. The interladder couplings can be classified into two distinct types according to their geometry. The coupling $J_{\perp}$ connects spins located on opposite legs of the neighboring ladders. In contrast, the coupling $J_{\parallel}$ links spins belonging to the same legs. The interladder interactions $J_{\perp}$ and $J_{\parallel}$ are significantly weaker than $J$ and $J'$. Therefore, Bi$_2$CuO$_3$(SO$_4$) should be deemed a quasi-1D magnet.

\subsection{Comparison to the experiment}
To facilitate a quantitative comparison between the exchange parameters obtained from DFT and the experimental data, we perform QMC simulations of the magnetic susceptibility. The inclusion of a weak diagonal interaction does not significantly affect the thermodynamic response of the ladder system~\cite{johnston2000}. Therefore, only the dominant rung and leg couplings were retained in the simulations. Since QMC simulations yield dimensionless quantities, the computed observables were rescaled using appropriate normalization factors to enable a direct comparison with the experimental results. To this end, dimensionless forms of the susceptibility, spin gap, and temperature were introduced:
\begin{equation}
\chi^{*} = \frac{\chi\, |J'|}{N_{\mathrm A}\,(g\,\mu_{\mathrm B})^{2}}, 
\qquad
\Delta^{*} = \frac{\Delta}{|J'|}, 
\qquad
t = \frac{k_{\mathrm B} T}{|J'|}
\end{equation}

All quantities were expressed in units of a characteristic energy scale, which was chosen as the absolute value of the rung coupling $|J'|$. Fixing this scale to unity, a series of simulations was performed for different ratios $J/|J'|$, allowing us to systematically explore the influence of the leg coupling strength.

Preliminary QMC simulations performed for the coupling ratios in the range of 1.0 $\le J/|J'| \le$ 2.5 reveal simple empirical linear relations between the ratios $J/|J'|$, $|J'|/J$, and the position of the susceptibility maximum, $T_{\max}$:
\begin{equation}
\begin{aligned}
\begin{aligned}
\frac{J}{|J'|} &= \frac{T_{\max}}{|J'|}m_1 + b_1, \\
\frac{|J'|}{J} &= \frac{T_{\max}}{J} m_2 + b_2,
\end{aligned}
\leftrightarrow
\begin{pmatrix}
1 & -b_1 \\
-b_2 & 1
\end{pmatrix}
\begin{pmatrix}
J \\
|J'|
\end{pmatrix}
=
T_{\max}
\begin{pmatrix}
m_1 \\
m_2
\end{pmatrix}
\end{aligned}
\notag
\end{equation}
where $m_1$ = 1.553, $b_1=-0.057$ and $m_2=-0.582$, $b_2$ = 1.294 are the corresponding linear regression parameters. The resulting system of equations enables a straightforward estimation of both exchange constants using only the experimentally determined position of the susceptibility maximum. From $T_{\max}$ $\approx 156$ K, the exchange couplings are estimated as $J$ $\approx 231$ K and $|J'|$ $\approx 208$ K, yielding $J/|J'|$ $\approx 1.24$. Since these values are subject to uncertainty inherent to the empirical parametrization, they were employed as initial estimates for the subsequent simulations of the full susceptibility curve.

The results obtained for 21 discrete values of the anisotropy ratios $J/$$|J'|$ within the interval [1.0, 1.31] were subsequently fitted using Pad\'e approximants following the procedure described in Ref.~\cite{johnston2000a}. This approach enables a smooth interpolation of the ratio $J/|J'|$ and allows for a continuous fitting of the experimental data. The experimental magnetic susceptibility was then modeled by introducing four additional free parameters, $\chi_0$, $C_{imp}$, $\theta_{\rm imp}$ and Landé $g$-factor, according to:
\begin{equation}
\chi_m = \chi_0 + \frac{C_{\mathrm{imp}}}{T - \theta_{\mathrm{imp}}}
+ \frac{N_{\mathrm A} g^2 \mu_{\mathrm B}^2}{|J'|}\,\chi^{*}
\end{equation}
The best fit of the experimental magnetic susceptibility in the whole temperature range returns $J=258$\,K, $J'=-208$\,K, $g=2.14$ together with $\chi_0$ = -0.35 $\times \, 10^{-4}$ emu\,mol$^{-1}$, $C_{\rm imp}$ = 9.10 $\times \, 10^{-3}$ emu\,K\,mol$^{-1}$, and $\theta_{\rm imp}$ = -7\,K. The interaction parameter of the impurity part is consistent with the analysis of the ESR signal that returned $\theta_{\rm imp}=-10$\,K (Sec.~\ref{sec:esr}), whereas the $C_{\rm imp}$ value corresponds to 2.4\% of the impurity spins.


The spin gap was extracted for each simulated susceptibility curve by fitting the low-temperature data using the expression proposed in Ref.~\cite{troyer1994}: $\chi^{*} = (A/\sqrt{t})\,\mathrm{e}^{-\Delta^{*}/t}$, where $\Delta^{*} = \Delta/ |J'|$. The resulting $\Delta^{*}$ values listed in Table S1~\cite{SM} were subsequently rescaled by the magnitude of the rung coupling $|J'|$. The spin gaps obtained for the investigated coupling ratios amount to approximately 10\,\% of the rung exchange interaction $J'$. Using the exchange parameters from the susceptibility fit, we find $\Delta\simeq 28$\,K for the isolated spin ladder in Bi$_2$CuO$_3$(SO$_4$).


\section{Discussion and Summary}

The magnetic properties of Bi$_2$CuO$_3$(SO$_4$) are successfully rationalized within a two-leg spin-ladder model. The diagonal couplings within the ladder are very weak, leading to only a minor frustration. An interesting feature of the material is its combination of the FM rung coupling and AFM leg coupling of comparable strength, in contrast to the majority of Cu-based spin ladders, which are purely AFM. The FM nature of the rung coupling leads to a rather small spin gap, less than 10\% of $J$. Therefore, materials with the FM-AFM spin ladders are more prone to developing long-range magnetic order even from weak interladder couplings. This is likely the case of Bi$_2$CuO$_3$(SO$_4$) where both magnetic susceptibility and heat capacity show a weak anomaly at $T_N=16$\,K. With $T_N/J$ as low as 0.06, most of the magnetic entropy is released above $T_N$, and the transition manifests itself only by a small kink in $C_p/T$ instead of a larger $\lambda$-type anomaly. This situation resembles some other spin-ladder compounds, such as Ba$_2$CuTeO$_6$ where  magnetic ordering leads to a tiny anomaly in the magnetic susceptibility with no clear signatures in the specific heat~\cite{rao2016,gibbs2017}.

It is further instructive to compare the coupling strength in various Cu-based spin-ladder compounds. Their rung couplings can be either short-range (mediated by a Cu-L-Cu bridge, where L is the ligand atom) or long-range (mediated by several ligand atoms), whereas leg couplings are typically long-range. Relevant materials and their exchange parameters are summarized in Table~\ref{tab:comparison}. One sees that FM rungs are rare and only occur in geometries with the short-range couplings. Even in this case, represented by the Cu$_2$O$_6$ dimers, the rung coupling can be either weakly FM or weakly AFM, as exemplified by IPA-CuCl$_3$ and PHCC, respectively. With its Cu--O--Cu angle of $90.1^{\circ}$, Bi$_2$CuO$_3$(SO$_4$) stands out as the spin-ladder material with the strong FM exchange on the rung. The magnitude of the FM exchange in this case is comparable to materials where CuO$_4$ plaquettes condense into chains with the strong FM nearest-neighbor coupling, such as $J_1=-228$\,K in Li$_2$CuO$_2$~\cite{lorenz2009}. 


\begin{table}[t]
\caption{\label{tab:comparison}
Comparison of the leg ($J_{leg}$) and rung ($J_{rung}$) exchange couplings in representative two-leg $S=\tfrac{1}{2}$ spin-ladder compounds. For each material, the dominant superexchange pathways (S for short-range and L for long-range) and the sign of the interactions (AFM or FM) are indicated. Exchange constants are given in K; negative values denote ferromagnetic couplings. For (CPA)$_2$CuBr$_4$, the exchange couplings are averaged over two inequivalent ladders reported in Ref.~\cite{philippe2025}. Experimental values (fits to the magnetic susceptibility and/or excitation spectra) are given in each case.  
}
\begin{ruledtabular}
\begin{tabular}{crrc}
compound & $J_{rung}$ (K) & $J_{leg}$ (K) & Ref. \medskip \\
\hline \medskip 
BPCB     & 13 (L) & $~\!3$ (L) & \cite{watson2001,savici2009}  \medskip \\
DLCB     & 7 (L)  &  7 (L)    & \cite{hong2014}  \medskip \\
(CPA)$_2$CuBr$_4$ & 10 (L) & 9 (L) & \cite{philippe2025} \medskip \\
IPA-CuCl$_3$ & $-27$ (S) &  14 (L)  &  \cite{fischer2011} \medskip \\  
PHCC         & 23 (S) & 14 (L) & \cite{tsirlin2025} \medskip\\
DIMPY        & 9 (L)  & 17 (L) &  \cite{hong2010,schmidiger2012}  \medskip \\
Ba$_2$CuTeO$_6$   & $\sim\!90$ (L) & $\sim\!90$ (L)  &    \cite{gibbs2017,glamazda2017} \medskip \\
BiCu$_2$PO$_6$ & $\sim\!110$ (L) &  $\sim\!110$ (L)  &  \cite{plumb2016,splinter2016} \medskip \\ 
KCu$_2$BiO$_2$(SO$_4$)$_2$ & 205 (L) & 117 (L) & \cite{shvanskaya2026} \medskip \\
Bi$_2$CuO$_3$(SO$_4$) & $-208$ (S) &  258 (L)  &  this work \\  
\end{tabular}
\end{ruledtabular}
\end{table}

The leg coupling in Bi$_2$CuO$_3$(SO$_4$) is equally remarkable. It largely exceeds the respective couplings in other spin-ladder magnets with long-range superexchange. In fact, $J=258$\,K manifests one of the strongest long-range superexchange couplings among all Cu$^{2+}$ materials reported to date. Whereas Cu--O--O--Cu pathways are known to be efficient because of the large O $2p$ contribution to the respective Wannier functions (Fig.~\ref{fig6}), the resulting exchange couplings rarely exceed 100\,K, given the large separation of at least 5\,\r A between the interacting Cu$^{2+}$ ions. The $J$ value in Bi$_2$CuO$_3$(SO$_4$) lies well beyond this typical range. A similarly strong coupling of about 260\,K has been reported in Na$_2$Cu$_2$TeO$_6$~\cite{gao2020,shangguan2021}. In both cases, the CuO$_4$ plaquettes are perfectly co-aligned, thanks to the bonding network created by Bi$^{3+}$ and Te$^{6+}$ forming short contacts to oxygen. While these ``observer'' atoms do not contribute any states near the Fermi level and do not directly participate in superexchange, they apparently play an important role in stabilizing the optimal geometry for the long-range superexchange coupling.

In summary, we reported magnetic properties and determined microscopic interaction parameters in the spin-ladder material Bi$_2$CuO$_3$(SO$_4$). The established spin-ladder model features  ferromagnetic rungs and antiferromagnetic legs. Both couplings largely exceed in magnitude magnetic interaction parameters in other spin-ladder materials with similar superexchange geometries. A weak anomaly is consistently observed around 16\,K and indicates a likely transition into a magnetically ordered state stabilized by non-negligible interladder couplings. 



\acknowledgments
Computations for this work were done (in part) using resources of the Leipzig University Computing Center. The work in Dresden was supported by the DFG through the W\"{u}rzburg-Dresden Cluster of Excellence on Complexity, Topology and Dynamics in Quantum Matter -- \textit{ctd.qmat} (EXC 2147, project No.\ 390858490) and the SFB 1143 (Project No.\ 247310070), as well as by HLD at HZDR, member of the European Magnetic Field Laboratory (EMFL). V.A.G. acknowledges the personal scholarship from DAAD under grant no. 91837667. R.A.R.H. was supported by a FES-Stipendium.  We acknowledge ESRF for providing the beamtime at ID22 and thank Andy Fitch, Catherine Dejoie, and Javier Gainza for their technical support during the measurement. 


\bibliography{BiCu-sulfate}

@article{haldane1983,
title = {Continuum dynamics of the 1-D Heisenberg antiferromagnet: Identification with the O(3) nonlinear sigma model},
Author = {F. D. M. Haldane},
journal = {Physics Letters A},
volume = {93},
number = {9},
pages = {464-468},
year = {1983},
issn = {0375-9601},
doi = {https://doi.org/10.1016/0375-9601(83)90631-X}
}

@article{dagotto1996,
title = {Surprises on the Way from One- to Two-Dimensional Quantum Magnets: The Ladder Materials},
author = {Elbio Dagotto and T. M. Rice},
journal = {Science},
volume = {271},
number = {5249},
pages = {618-623},
year = {1996},
doi = {10.1126/science.271.5249.618}
}

@article{lemmens2003,
title = {Magnetic light scattering in low-dimensional quantum spin systems},
author = {P. Lemmens, G. G\"untherodt, C. Gros},
journal = {Physics Reports},
volume = {375},
number = {1},
pages = {1-103},
year = {2003},
issn = {0370-1573},
doi = {https://doi.org/10.1016/S0370-1573(02)00321-6}
}

@Article{lebernegg2017,
  Title                    = {Frustrated spin chain physics near the {Majumdar-Ghosh} point in szenicsite {Cu$_3$(MoO$_4$)(OH)$_4$}},
  Author                   = {S. Lebernegg and O. Janson and I. Rousochatzakis and S. Nishimoto and H. Rosner and A. A. Tsirlin},
  Journal                  = {Phys. Rev. B},
  Year                     = {2017},
  Pages                    = {035145},
  Volume                   = {95},
  Doi                      = {10.1103/PhysRevB.95.035145}
}

@Article{kulbakov2022a,
  title   = {Coupled frustrated ferromagnetic and antiferromagnetic quantum spin chains in the quasi-one-dimensional mineral antlerite {Cu$_3$SO$_4$(OH)$_4$}},
  author  = {Kulbakov, Anton A. and Kononenko, Denys Y. and Nishimoto, Satoshi and Stahl, Quirin and Chakkingal, Aswathi Mannathanath and Feig, Manuel and Gumeniuk, Roman and Skourski, Yurii and Bhaskaran, Lakshmi and Zvyagin, Sergei A. and Embs, Jan Peter and Puente-Orench, In\'es and Wildes, Andrew and Geck, Jochen and Janson, Oleg and Inosov, Dmytro S. and Peets, Darren C.},
  journal = {Phys. Rev. B},
  year    = {2022},
  volume  = {106},
  pages   = {L020405},
  doi     = {10.1103/PhysRevB.106.L020405}
}

@article{glamazda2017,
  author = {Glamazda, A. and Choi, Y. S. and Do, S.-H. and Lee, S. and Lemmens, P. and Ponomaryov, A. N. and Zvyagin, S. A. and Wosnitza, J. and Sari, Dita Puspita and Watanabe, I. and Choi, K.-Y.},
  title = {Quantum criticality in the coupled two-leg spin ladder {Ba$_2$CuTeO$_6$}},
  journal = {Phys. Rev. B},
  volume = {95},
  issue = {18},
  pages = {184430},
  numpages = {9},
  year = {2017},
  month = {May},
  publisher = {American Physical Society},
  doi = {10.1103/PhysRevB.95.184430},
}

@article{tsirlin2010,
  author = {Tsirlin, Alexander A. and Rousochatzakis, Ioannis and Kasinathan, Deepa and Janson, Oleg and Nath, Ramesh and Weickert, Franziska and Geibel, Christoph and L\"auchli, Andreas M. and Rosner, Helge},
  title = {Bridging frustrated-spin-chain and spin-ladder physics: Quasi-one-dimensional magnetism of {BiCu$_2$PO$_6$}},
  journal = {Phys. Rev. B},
  volume = {82},
  issue = {14},
  pages = {144426},
  numpages = {15},
  year = {2010},
  month = {Oct},
  publisher = {American Physical Society},
  doi = {10.1103/PhysRevB.82.144426},
}

@article{azuma1994,
  title = {Observation of a Spin Gap in {SrCu$_2$O$_3$} Comprising Spin-\textonehalf{} Quasi-1$D$ Two-Leg Ladders},
  author = {Azuma, M. and Hiroi, Z. and Takano, M. and Ishida, K. and Kitaoka, Y.},
  journal = {Phys. Rev. Lett.},
  volume = {73},
  issue = {25},
  pages = {3463--3466},
  numpages = {0},
  year = {1994},
  month = {Dec},
  publisher = {American Physical Society},
  doi = {10.1103/PhysRevLett.73.3463},
  url = {https://link.aps.org/doi/10.1103/PhysRevLett.73.3463}
}

@article{koteswararao2007,
  title = {Spin-gap behavior in the two-leg spin-ladder {BiCu$_2$PO$_6$}},
  author = {Koteswararao, B. and Salunke, S. and Mahajan, A. V. and Dasgupta, I. and Bobroff, J.},
  journal = {Phys. Rev. B},
  volume = {76},
  issue = {5},
  pages = {052402},
  numpages = {4},
  year = {2007},
  month = {Aug},
  publisher = {American Physical Society},
  doi = {10.1103/PhysRevB.76.052402},
  url = {https://link.aps.org/doi/10.1103/PhysRevB.76.052402}
}

@article{pilch2025,
  title = {Low-energy spin excitations in field-induced phases of the spin-ladder antiferromagnet {BiCu$_2$PO$_6$}},
  author = {Pilch, Patrick and Amelin, Kirill and Schmiedinghoff, Gary and Reinold, Anneke and Zhu, Changqing and Povarov, Kirill Yu. and Zvyagin, Sergei and Engelkamp, Hans and Lan, Yin-Ping and Shu, Guo-Jiun and Chou, F. C. and Nagel, Urmas and R\~o\~om, Toomas and Uhrig, G\"otz S. and Fauseweh, Benedikt and Wang, Zhe},
  journal = {Phys. Rev. B},
  volume = {111},
  issue = {2},
  pages = {024423},
  numpages = {12},
  year = {2025},
  month = {Jan},
  publisher = {American Physical Society},
  doi = {10.1103/PhysRevB.111.024423},
}

@article{kohama2012,
  title = {Anisotropic cascade of field-induced
phase transitions in the frustrated spin-ladder system
{${\text{BiCu}}_{2}{\text{PO}}_{6}$}},
  author = {Y. Kohama and S. Wang and A. Uchida and K. Prsa and S. Zvyagin and Y.
Skourski and R. D. McDonald and L. Balicas and H. M. R{\o}nnow and C. R\"uegg and M. Jaime},
  journal = {Phys. Rev. Lett.},
  volume = {109},
  issue = {16},
  pages = {167204},
  numpages = {5},
  year = {2012},
  month = {Oct},
  publisher = {American Physical Society},
  doi = {10.1103/PhysRevLett.109.167204},
}

@article{choi2013,
  title = {Evidence for Dimer Crystal Melting in the Frustrated Spin-Ladder System {BiCu$_2$PO$_6$}},
  author = {Choi, K.-Y. and Hwang, J. W. and Lemmens, P. and Wulferding, D. and Shu, G. J. and Chou, F. C.},
  journal = {Phys. Rev. Lett.},
  volume = {110},
  issue = {11},
  pages = {117204},
  numpages = {6},
  year = {2013},
  month = {Mar},
  publisher = {American Physical Society},
  doi = {10.1103/PhysRevLett.110.117204},
}

@ARTICLE{pbe96,
  title = {Generalized Gradient Approximation Made Simple},
  author = {Perdew, J. P. and Burke, K. and Ernzerhof, M.},
  journal = {Phys. Rev. Lett.},
  volume = {77},
  issue = {18},
  pages = {3865},
  year = {1996},
  doi = {10.1103/PhysRevLett.77.3865},
}

@ARTICLE{fplo,
  title = {Full-potential nonorthogonal local-orbital minimum-basis band-structure scheme},
  author = {Koepernik, K. and Eschrig, H.},
  journal = {Phys. Rev. B},
  volume = {59},
  issue = {3},
  pages = {1743},
  numpages = {0},
  year = {1999},
  doi = {10.1103/PhysRevB.59.1743}
}

@ARTICLE{mazurenko2014,
  title = {Nonfrustrated Interlayer Order and its Relevance to the Bose-Einstein Condensation of Magnons in {BaCuSi$_2$O$_6$}},
  author = {Mazurenko, Vladimir V. and Valentyuk, Maria V. and Stern, Raivo and Tsirlin, Alexander A.},
  journal = {Phys. Rev. Lett.},
  volume = {112},
  issue = {10},
  pages = {107202},
  numpages = {5},
  year = {2014},
  month = {Mar},
  publisher = {American Physical Society},
  doi = {10.1103/PhysRevLett.112.107202},
  url = {https://link.aps.org/doi/10.1103/PhysRevLett.112.107202}
}

@article{bag2021,
  title = {Low-dimensional magnetism of {BaCuTe$_2$O$_6$}},
  author = {Bag, P. and Ahmed, N. and Singh, Vikram and Sahoo, M. and Tsirlin, A. A. and Nath, R.},
  journal = {Phys. Rev. B},
  volume = {103},
  issue = {13},
  pages = {134410},
  numpages = {10},
  year = {2021},
  month = {Apr},
  publisher = {American Physical Society},
  doi = {10.1103/PhysRevB.103.134410},
  url = {https://link.aps.org/doi/10.1103/PhysRevB.103.134410}
}

@article{xiang2011,
  title = {Predicting the spin-lattice order of frustrated systems from first principles},
  author = {Xiang, H. J. and Kan, E. J. and Wei, Su-Huai and Whangbo, M.-H. and Gong, X. G.},
  journal = {Phys. Rev. B},
  volume = {84},
  issue = {22},
  pages = {224429},
  numpages = {5},
  year = {2011},
  month = {Dec},
  publisher = {American Physical Society},
  doi = {10.1103/PhysRevB.84.224429},
  url = {https://link.aps.org/doi/10.1103/PhysRevB.84.224429}
}

@article{albuquerque2007,
title = {The {ALPS} project release 1.3: Open-source software for strongly correlated systems},
journal = {J. Magn. Magn. Mater.},
volume = {310},
pages = {1187-1193},
year = {2007},
doi = {10.1016/j.jmmm.2006.10.304},
author = {A.F. Albuquerque and F. Alet and P. Corboz and P. Dayal and A. Feiguin and S. Fuchs and L. Gamper and E. Gull and S. Gürtler and A. Honecker and R. Igarashi and M. Körner and A. Kozhevnikov and A. Läuchli and S.R. Manmana and M. Matsumoto and I.P. McCulloch and F. Michel and R.M. Noack and G. Pawłowski and L. Pollet and T. Pruschke and U. Schollwöck and S. Todo and S. Trebst and M. Troyer and P. Werner and S. Wessel},
}

@misc{johnston2000,
      title={Magnetic Susceptibilities of Spin-1/2 Antiferromagnetic {Heisenberg} Ladders and Applications to Ladder Oxide Compounds}, 
      author={D. C. Johnston and M. Troyer and S. Miyahara and D. Lidsky and K. Ueda and M. Azuma and Z. Hiroi and M. Takano and M. Isobe and Y. Ueda and M. A. Korotin and V. I. Anisimov and A. V. Mahajan and L. L. Miller},
      eprint={cond-mat/0001147},
      year = {2000},
      archivePrefix={arXiv}
}

@article{tsirlin2010a,
  title = {{$\beta$-Cu$_2$V$_2$O$_7$}: A spin-$\frac{1}{2}$ honeycomb lattice system},
  author = {Tsirlin, Alexander A. and Janson, Oleg and Rosner, Helge},
  journal = {Phys. Rev. B},
  volume = {82},
  issue = {14},
  pages = {144416},
  numpages = {10},
  year = {2010},
  month = {Oct},
  publisher = {American Physical Society},
  doi = {10.1103/PhysRevB.82.144416},
  url = {https://link.aps.org/doi/10.1103/PhysRevB.82.144416}
}

@article{mazurenko2007,
  title = {Wannier functions and exchange integrals: The example of {LiCu$_2$O$_2$}},
  author = {Mazurenko, V. V. and Skornyakov, S. L. and Kozhevnikov, A. V. and Mila, F. and Anisimov, V. I.},
  journal = {Phys. Rev. B},
  volume = {75},
  issue = {22},
  pages = {224408},
  numpages = {7},
  year = {2007},
  month = {Jun},
  publisher = {American Physical Society},
  doi = {10.1103/PhysRevB.75.224408},
  url = {https://link.aps.org/doi/10.1103/PhysRevB.75.224408}
}

@article{tsirlin2010b,
  title = {Uniform spin-chain physics arising from {N}---{C}---{N} bridges in $\text{CuNCN}$, the nitride analog of the copper oxides},
  author = {Tsirlin, Alexander A. and Rosner, Helge},
  journal = {Phys. Rev. B},
  volume = {81},
  issue = {2},
  pages = {024424},
  numpages = {10},
  year = {2010},
  month = {Jan},
  publisher = {American Physical Society},
  doi = {10.1103/PhysRevB.81.024424},
  url = {https://link.aps.org/doi/10.1103/PhysRevB.81.024424}
}

@article{tsirlin2011,
  title = {Spiral ground state in the quasi-two-dimensional spin-$\frac{1}{2}$ system {Cu$_2$GeO$_4$}},
  author = {Tsirlin, Alexander A. and Zinke, Ronald and Richter, Johannes and Rosner, Helge},
  journal = {Phys. Rev. B},
  volume = {83},
  issue = {10},
  pages = {104415},
  numpages = {7},
  year = {2011},
  month = {Mar},
  publisher = {American Physical Society},
  doi = {10.1103/PhysRevB.83.104415},
  url = {https://link.aps.org/doi/10.1103/PhysRevB.83.104415}
}

@article{eschrig2009,
  title = {Tight-binding models for the iron-based superconductors},
  author = {Eschrig, Helmut and Koepernik, Klaus},
  journal = {Phys. Rev. B},
  volume = {80},
  issue = {10},
  pages = {104503},
  numpages = {14},
  year = {2009},
  month = {Sep},
  publisher = {American Physical Society},
  doi = {10.1103/PhysRevB.80.104503},
  url = {https://link.aps.org/doi/10.1103/PhysRevB.80.104503}
}

@article{johnston2000a,
  title = {Thermodynamics of spin $S=1/2$ antiferromagnetic uniform and alternating-exchange Heisenberg chains},
  author = {Johnston, D. C. and Kremer, R. K. and Troyer, M. and Wang, X. and Kl\"umper, A. and Bud'ko, S. L. and Panchula, A. F. and Canfield, P. C.},
  journal = {Phys. Rev. B},
  volume = {61},
  issue = {14},
  pages = {9558--9606},
  numpages = {0},
  year = {2000},
  month = {Apr},
  publisher = {American Physical Society},
  doi = {10.1103/PhysRevB.61.9558},
  url = {https://link.aps.org/doi/10.1103/PhysRevB.61.9558}
}

@article{troyer1994,
  title = {Thermodynamics and spin gap of the Heisenberg ladder calculated by the look-ahead Lanczos algorithm},
  author = {Troyer, Matthias and Tsunetsugu, Hirokazu and W\"urtz, Diethelm},
  journal = {Phys. Rev. B},
  volume = {50},
  issue = {18},
  pages = {13515--13527},
  numpages = {0},
  year = {1994},
  month = {Nov},
  publisher = {American Physical Society},
  doi = {10.1103/PhysRevB.50.13515},
  url = {https://link.aps.org/doi/10.1103/PhysRevB.50.13515}
}

@article{schmidiger2012,
  title = {Spectral and Thermodynamic Properties of a Strong-Leg Quantum Spin Ladder},
  author = {Schmidiger, D. and Bouillot, P. and M\"uhlbauer, S. and Gvasaliya, S. and Kollath, C. and Giamarchi, T. and Zheludev, A.},
  journal = {Phys. Rev. Lett.},
  volume = {108},
  issue = {16},
  pages = {167201},
  numpages = {5},
  year = {2012},
  month = {Apr},
  publisher = {American Physical Society},
  doi = {10.1103/PhysRevLett.108.167201},
  url = {https://link.aps.org/doi/10.1103/PhysRevLett.108.167201}
}

@article{hong2014,
  title = {Magnetic ordering induced by interladder coupling in the spin-$\frac{1}{2}$ Heisenberg two-leg ladder antiferromagnet {C$_9$H$_{18}$N$_2$CuBr$_4$}},
  author = {Hong, Tao and Schmidt, K. P. and Coester, K. and Awwadi, F. F. and Turnbull, M. M. and Qiu, Y. and Rodriguez-Rivera, J. A. and Zhu, M. and Ke, X. and Aoyama, C. P. and Takano, Y. and Cao, Huibo and Tian, W. and Ma, J. and Custelcean, R. and Zhou, H. D. and Matsuda, M.},
  journal = {Phys. Rev. B},
  volume = {89},
  issue = {17},
  pages = {174432},
  numpages = {6},
  year = {2014},
  month = {May},
  publisher = {American Physical Society},
  doi = {10.1103/PhysRevB.89.174432},
  url = {https://link.aps.org/doi/10.1103/PhysRevB.89.174432}
}

@article{lorenz2009,
doi = {10.1209/0295-5075/88/37002},
year = {2009},
volume = {88},
pages = {37002},
author = {Lorenz, W. E. A. and Kuzian, R. O. and Drechsler, S.-L. and Stein, W.-D. and Wizent, N. and Behr, G. and M\'alek, J. and Nitzsche, U. and Rosner, H. and Hiess, A. and Schmidt, W. and Klingeler, R. and Loewenhaupt, M. and Büchner, B.},
title = {Highly dispersive spin excitations in the chain cuprate {Li$_2$CuO$_2$}},
journal = {Europhys. Lett.}
}

@article{gao2020,
  title = {Weakly coupled alternating $S=\frac{1}{2}$ chains in the distorted honeycomb lattice compound {Na$_2$Cu$_2$TeO$_6$}},
  author = {Gao, Shang and Lin, Ling-Fang and May, Andrew F. and Rai, Binod K. and Zhang, Qiang and Dagotto, Elbio and Christianson, Andrew D. and Stone, Matthew B.},
  journal = {Phys. Rev. B},
  volume = {102},
  issue = {22},
  pages = {220402},
  numpages = {6},
  year = {2020},
  month = {Dec},
  publisher = {American Physical Society},
  doi = {10.1103/PhysRevB.102.220402},
  url = {https://link.aps.org/doi/10.1103/PhysRevB.102.220402}
}

@Article{shangguan2021,
  author  = {Shangguan, Yanyan and Bao, Song and Dong, Zhao-Yang and Cai, Zhengwei and Wang, Wei and Huang, Zhentao and Ma, Zhen and Liao, Junbo and Zhao, Xiaoxue and Kajimoto, Ryoichi and Iida, Kazuki and Voneshen, David and Yu, Shun-Li and Li, Jian-Xin and Wen, Jinsheng},
  journal = {Phys. Rev. B},
  title   = {Evidence for strong correlations at finite temperatures in the dimerized magnet {Na$_2$Cu$_2$TeO$_6$}},
  year    = {2021},
  pages   = {224430},
  volume  = {104},
  doi     = {10.1103/PhysRevB.104.224430}
}

@article{rao2016,
  title = {Tellurium-bridged two-leg spin ladder in {Ba$_2$CuTeO$_6$}},
  author = {Rao, G. Narsinga and Sankar, R. and Singh, Akansha and Muthuselvam, I. Panneer and Chen, W. T. and Singh, Viveka Nand and Guo, Guang-Yu and Chou, F. C.},
  journal = {Phys. Rev. B},
  volume = {93},
  issue = {10},
  pages = {104401},
  numpages = {10},
  year = {2016},
  month = {Mar},
  publisher = {American Physical Society},
  doi = {10.1103/PhysRevB.93.104401},
  url = {https://link.aps.org/doi/10.1103/PhysRevB.93.104401}
}

@article{gibbs2017,
  title = {$S=\frac{1}{2}$ quantum critical spin ladders produced by orbital ordering in {Ba$_2$CuTeO$_6$}},
  author = {Gibbs, A. S. and Yamamoto, A. and Yaresko, A. N. and Knight, K. S. and Yasuoka, H. and Majumder, M. and Baenitz, M. and Saines, P. J. and Hester, J. R. and Hashizume, D. and Kondo, A. and Kindo, K. and Takagi, H.},
  journal = {Phys. Rev. B},
  volume = {95},
  issue = {10},
  pages = {104428},
  numpages = {6},
  year = {2017},
  month = {Mar},
  publisher = {American Physical Society},
  doi = {10.1103/PhysRevB.95.104428},
  url = {https://link.aps.org/doi/10.1103/PhysRevB.95.104428}
}

@article{savici2009,
  title = {Neutron scattering evidence for isolated spin-$\frac{1}{2}$ ladders in {(C$_5$D$_{12}$N)$_2$CuBr$_4$}},
  author = {Savici, A. T. and Granroth, G. E. and Broholm, C. L. and Pajerowski, D. M. and Brown, C. M. and Talham, D. R. and Meisel, M. W. and Schmidt, K. P. and Uhrig, G. S. and Nagler, S. E.},
  journal = {Phys. Rev. B},
  volume = {80},
  issue = {9},
  pages = {094411},
  numpages = {8},
  year = {2009},
  month = {Sep},
  publisher = {American Physical Society},
  doi = {10.1103/PhysRevB.80.094411},
  url = {https://link.aps.org/doi/10.1103/PhysRevB.80.094411}
}

@article{fischer2011,
doi = {10.1209/0295-5075/96/47001},
url = {https://doi.org/10.1209/0295-5075/96/47001},
year = {2011},
month = {nov},
publisher = {},
volume = {96},
number = {4},
pages = {47001},
author = {Fischer, T. and Duffe, S. and Uhrig, G. S.},
title = {Microscopic model for Bose-Einstein condensation and quasiparticle decay},
journal = {Europhysics Letters},
}

@article{lu2014,
  author       = {L{\"u}, M. and Colmont, M. and Kabbour, H. and Colis, S. and Mentr{\'e}, O.},
  title        = {Revised {Bi/M} layered oxo-sulfate ({M} = {Co}, {Cu}): a structural and magnetic study},
  journal      = {Inorg. Chem.},
  year         = {2014},
  volume       = {53},
  number       = {13},
  pages        = {6969--6978},
  doi          = {10.1021/ic500877z}
}

@Article{siidra2020,
  Title                    = {Expanding Family of Litharge-Derived Sulfate Minerals and Synthetic Compounds: Preparation and
Crystal Structures of {[Bi$_2$CuO$_3$]SO$_4$} and {[Ln$_2$O$_2$]SO$_4$} ({Ln = Dy and Ho})},
  Author                   = {O. Siidra and D. Charkin and I. Plokhikh and E Nazarchuk and A. Holzheid and G. Akimov},
  Journal                  = {Minerals},
  Year                     = {2020},
  Pages                    = {887},
  Volume                   = {10},
  Doi                      = {10.3390/min10100887}
}

@Article{inosov2018,
  Title                    = {Quantum magnetism in minerals},
  Author                   = {D. S. Inosov},
  Journal                  = {Adv. Phys.},
  Year                     = {2018},
  Pages                    = {149-252},
  Volume                   = {67},
  Doi                      = {10.1080/00018732.2018.1571986}
}

@Article{zhang2020,
  Title                    = {Coexistence and Interaction of Spinons and Magnons in an Antiferromagnet with Alternating Antiferromagnetic and Ferromagnetic Quantum Spin Chains},
  Author                   = {Zhang, H. and Zhao, Z. and Gautreau, D. and Raczkowski, M. and Saha, A. and Garlea, V. O. and Cao, H. and Hong, T. and Jeschke, H. O. and Mahanti, Subhendra D. and Birol, T. and Assaad, F. F. and Ke, X.},
  Journal                  = {Phys. Rev. Lett.},
  Year                     = {2020},
  Pages                    = {037204},
  Volume                   = {125},
  Doi                      = {10.1103/PhysRevLett.125.037204}
}

@Article{fitch2023,
  author  = {A. Fitch and C. Dejoie and E. Covacci and G. Confalonieri and O. Grendal and L. Claustre and P. Guillou and J. Kieffer and W. {de Nolf} and S. Petitdemange and M. Ruat and Y. Watier},
  title   = {{ID22} -- the high-resolution powder-diffraction beamline at {ESRF}},
  journal = {J. Synch. Rad.},
  year    = {2023},
  volume  = {30},
	pages   = {1003-1012},
  doi     = {10.1107/S1600577523004915}
}

@Article{jana2006,
  Title                    = {Crystallographic Computing System {JANA2006}: General features},
  Author                   = {Pet{\u r}{\'\i}{\u c}ek, V. and Du{\u s}ek, M. and Palatinus, L.},
  Journal                  = {Z. Krist.},
  Year                     = {2014},
  Pages                    = {345--352},
  Volume                   = {229},
  Doi	                     = {10.1515/zkri-2014-1737}
}

@Article{sugai1999,
  Title                    = {Anisotropic exchange integrals in the two-leg spin ladder {LaCuO$_{2.5}$}},
  Author                   = {S. Sugai and T. Shinoda and N. Kobayashi and Z. Hiroi and M. Takano},
  Journal                  = {Phys. Rev. B},
  Year                     = {1999},
  Pages                    = {R6969--R6972},
  Volume                   = {60},
  Doi                      = {10.1103/PhysRevB.60.R6969}
}

@Article{kiryukhin2001,
  Title                    = {Magnetic properties of the {$S=1/2$} quasi-one-dimensional antiferromagnet {CaCu$_2$O$_3$}},
  Author                   = {V. Kiryukhin and Y. J. Kim and K. J. Thomas and F. C. Chou and R. W. Erwin and Q. Huang and M. A. Kastner and R. J. Birgeneau},
  Journal                  = {Phys. Rev. B},
  Year                     = {2001},
  Number                   = {14},
  Pages                    = {144418},
  Volume                   = {63},
  Doi                      = {10.1103/PhysRevB.63.144418}
}

@Article{mentre2009,
  Title                    = {Incommensurate spin correlation driven by frustration in {BiCu$_2$PO$_6$}},
  Author                   = {O. Mentr\'e and E. Janod and P. Rabu and M. Hennion and F. Leclercq-Hugeux and J. Kang and C. Lee and M.-H. Whangbo and S. Petit},
  Journal                  = {Phys. Rev. B},
  Year                     = {2009},
  Pages                    = {180413(R)},
  Volume                   = {80},
  Doi                      = {10.1103/PhysRevB.80.180413}
}

@Article{plumb2016,
  Title                    = {Quasiparticle-continuum level repulsion in a quantum magnet},
  Author                   = {K. W. Plumb and K. Hwang and Y. Qiu and L. W. Harriger and G. E. Granroth and A. I. Kolesnikov and G. J. Shu and F. C. Chou and Ch. R\"uegg and Y. B. Kim and Y.-J. Kim},
  Journal                  = {Nature Phys.},
  Year                     = {2016},
  Pages                    = {224-229},
  Volume                   = {12},
  Doi                      = {10.1038/nphys3566}
}

@Article{casola2013,
  Title                    = {Field-Induced Quantum Soliton Lattice in a Frustrated Two-Leg Spin-1/2 Ladder},
  Author                   = {F. Casola and T. Shiroka and A. Feiguin and S. Wang and M. S. Grbi{\'c} and M. Horvati{\'c} and S. Kr\"amer and S. Mukhopadhyay and K. Conder and C. Berthier and H.-R. Ott and H. M. R{\o}nnow and Ch. R\"uegg and J. Mesot},
  Journal                  = {Phys. Rev. Lett.},
  Year                     = {2013},
  Pages                    = {187201},
  Volume                   = {110},
  Doi                      = {10.1103/PhysRevLett.110.187201}
}

@Article{splinter2016,
  Title                    = {Minimal model for the frustrated spin ladder system {BiCu$_2$PO$_6$}},
  Author                   = {L. Splinter and N. A. Drescher and H. Krull and G. S. Uhrig},
  Journal                  = {Phys. Rev. B},
  Year                     = {2016},
  Pages                    = {155115},
  Volume                   = {94},
  Doi                      = {10.1103/PhysRevB.94.155115}
}

@Article{hwang2016,
  Title                    = {Theory of triplon dynamics in the quantum magnet {BiCu$_2$PO$_6$}},
  Author                   = {K. Hwang and Y. B. Kim},
  Journal                  = {Phys. Rev. B},
  Year                     = {2016},
  Pages                    = {235130},
  Volume                   = {93},
  Doi                      = {10.1103/PhysRevB.93.235130}
}

@article{loop,
  title = {Cluster Algorithms for General-{$\mathit{S}$} Quantum Spin Systems},
  author = {Todo, S. and Kato, K.},
  journal = {Phys. Rev. Lett.},
  volume = {87},
  pages = {047203},
  numpages = {4},
  year = {2001},
  doi = {10.1103/PhysRevLett.87.047203}
}

@article{aurivillius1987,
  title = {Pyrolysis Products of {Bi$_2$(SO$_4)_3$}. Crystal Structures of {Bi$_{26}$O$_{27}$(SO$_4)_{12}$} and {Bi$_{14}$O$_{16}$(SO$_4)_5$}},
  author = {Aurivillius, B.},
  journal = {Acta Chem. Scand.},
  volume = {A41},
  pages = {415-422},
  year = {1987},
  doi = {10.3891/acta.chem.scand.41a-0415}
}

@article{Zvyagin_PhysB_2004_ESRinCuGeO3,
title = {{High-field ESR study of the dimerized-incommensurate phase transition in the spin-Peierls compound CuGeO$_3$}},
journal = {Physica B},
volume = {346-347},
pages = {1},
year = {2004},
doi = {https://doi.org/10.1016/j.physb.2004.01.009},
url = {https://www.sciencedirect.com/science/article/pii/S0921452604000237},
author = {Zvyagin, S. A.  and Krzystek,  J. and {van Loosdrecht},  P. H. M. and Dhalenne,  G. and Revcolevschi,  A.}
}

@Article{sigrist1996,
  author  = {Manfred Sigrist and Akira Furusaki},
  title   = {Low-Temperature Properties of the Randomly Depleted {Heisenberg} Ladder},
  journal = {J. Phys. Soc. Jpn.},
  year    = {1997},
  volume  = {65},
  pages   = {2385},
  doi     = {10.1143/JPSJ.65.2385}
}

@Article{mikeska1997,
  author  = {Mikeska, H.-J. and Neugebauer, U. and Schollw\"ock, U.},
  title   = {Spin ladders with nonmagnetic impurities},
  journal = {Phys. Rev. B},
  year    = {1997},
  volume  = {55},
  pages   = {2955},
  doi     = {10.1103/PhysRevB.55.2955}
}

@Article{schmidiger2016,
  Title                    = {Emergent Interacting Spin Islands in a Depleted Strong-Leg {Heisenberg} Ladder},
  Author                   = {D. Schmidiger and K. Yu. Povarov and S. Galeski and N. Reynolds and R. Bewley and T. Guidi and J. Ollivier and A. Zheludev},
  Journal                  = {Phys. Rev. Lett.},
  Year                     = {2016},
  Pages                    = {257203},
  Volume                   = {116},
  Doi                      = {10.1103/PhysRevLett.116.257203}
}

@Article{watson2001,
  Title                    = {Magnetic spin ladder {(C$_5$H$_{12}$N)$_2$CuBr$_4$}: High-field magnetization and scaling near quantum criticality},
  Author                   = {B. C. Watson and V. N. Kotov and M. W. Meisel and D. W. Hall and G. E. Granroth and W. T. Montfrooij and S. E. Nagler and D. A Jensen and R. Backov and M. A. Petruska and G. E. Fanucci and D. R. Talham},
  Journal                  = {Phys. Rev. Lett.},
  Year                     = {2001},
  Number                   = {22},
  Pages                    = {5168--5171},
  Volume                   = {86},
  Doi                      = {10.1103/PhysRevLett.86.5168}
}

@Article{hong2010,
  Title                    = {Field-induced Tomonaga-Luttinger liquid phase of a two-leg spin-1/2 ladder with strong leg interactions},
  Author                   = {T. Hong and Y. H. Kim and C. Hotta and Y. Takano and G. Tremelling and M. M. Turnbull and C. P. Landee and H.-J. Kang and N. B. Christensen and K. Lefmann and K. P. Schmidt and G. S. Uhrig and C. Broholm},
  Journal                  = {Phys. Rev. Lett.},
  Year                     = {2010},
  Pages                    = {137207},
  Volume                   = {105},
  Doi                      = {10.1103/PhysRevLett.105.137207}
}

@Misc{SM,
  note        = {See Supplemental Material for the additional details of the crystal structure, comparison between the different batches, and for the simulated magnetization curves.}
}

@Misc{esrf,
  author = {Ginga, V. A. and Tsirlin, A. A.},
  title = {High-resolution x-ray diffraction study of the spin-ladder magnet {Bi$_2$CuO$_3$(SO$_4$)}},
  doi = {10.15151/ESRF-DC-2365344918},
  year = {2025}
}

@Misc{tsirlin2025,
  author        = {A. A. Tsirlin and O. Janson and I. Rousochatzakis},
  title         = {One-dimensional physics of the frustrated quantum magnet {PHCC}},
  archiveprefix = {arXiv},
  year = {2025},
  eprint        = {2512.17406}
}

@misc{philippe2025,
  author = {Philippe, J. and Elson, F. and Arh, T. and Sanz, S. and Metzelaars, M. and Tam, D. W. and Forslund, O. K. and Shliakhtun, O. and Jiang, C. and Lass, J. and Le, M. D. and Ollivier, J. and Bouillot, P. and Giamarchi, T. and Bartkowiak, M. and Mazzone, D. G. and K\"ogerler, P. and M{\aa}nsson, M. and L\"auchli, A. M. and Sassa, Y. and Janoschek, M. and Normand, B. and Simutis, G.},
  title = {Magnetic and phononic dynamics in the two-ladder quantum magnet {(C$_5$H$_9$NH$_3$)$_2$CuBr$_4$}},
  year = {2025},
  eprint = {2510.24556},
  archivePrefix = {arXiv},
  primaryClass = {cond-mat.str-el},
  url = {https://arxiv.org/abs/2510.24556}
}

@article{shvanskaya2026,
  author = {Shvanskaya, Larisa V. and Bushneva, Tatiana D. and Chareev, Dmitry A. and Maksimov, Pavel A. and Sudakov, Alexander A. and Kornienko, Ekaterina I. and Ushakov, Alexey V. and Streltsov, Sergey V. and Vasiliev, Alexander N.},
  title = {Peculiar Crystal Structure and Strong-Rung Spin Ladders in {KCu$_2$BiO$_2$(SO$_4$)$_2$}},
  journal = {Inorganic Chemistry},
  year = {2026},
  volume = {65},
  pages = {395--402},
  doi = {10.1021/acs.inorgchem.5c04485}
}

\end{document}